\documentclass[letterpaper]{article} 

\usepackage{xcolor}

\usepackage{aaai2026}  

\nocopyright

\usepackage{times}  
\usepackage{helvet}  
\usepackage{courier}  
\usepackage[hyphens]{url}  
\usepackage{graphicx} 
\usepackage{natbib}  
\usepackage{caption} 
\usepackage{xcolor}
\usepackage{algorithm}
\usepackage{algorithmic}

\usepackage[T1]{fontenc}

\usepackage{newfloat}
\usepackage{listings}

\DeclareCaptionStyle{rulednotop}{
    labelfont=normalfont,
    labelsep=colon,
    strut=off
}

\makeatletter
\newcommand{\fs@rulednotop}{%
    \def\@fs@cfont{\bfseries}%
    \let\@fs@capt\floatc@ruled
    \def\@fs@pre{}
    \def\@fs@mid{\kern2pt\hrule\kern2pt}%
    \def\@fs@post{\kern2pt\hrule\relax}%
    \let\@fs@iftopcapt\iftrue
}
\makeatother
\floatstyle{rulednotop}
\newfloat{listing}{tb}{lst}{}
\floatname{listing}{Listing}

\usepackage{tabularx}
\usepackage{tabularray}
\UseTblrLibrary{booktabs}
\usepackage{subcaption}
\usepackage{multirow}
\usepackage{booktabs}
\usepackage{adjustbox}

\usepackage{bm}

\usepackage[dvipsnames]{xcolor}

\usepackage{pgf}
\usepackage{amsmath}

\newcommand{\fscore}[2]{%
  \pgfmathparse{100*(#1)}%
  \pgfmathprintnumber[fixed,precision=2]{\pgfmathresult}
  $\pm$
  \pgfmathparse{100*(#2)}%
  \pgfmathprintnumber[fixed,precision=2]{\pgfmathresult}
}

\newcommand{\bfscore}[2]{%
  \pgfmathparse{100*(#1)}%
  $\mathbf{\pgfmathprintnumber[fixed,precision=2,assume math mode=true]{\pgfmathresult}}%
  \boldsymbol{\pm}%
  \pgfmathparse{100*(#2)}%
  \mathbf{\pgfmathprintnumber[fixed,precision=2,assume math mode=true]{\pgfmathresult}}$
}

\usepackage{xcolor}
\usepackage{dashrule}

\newcommand{\promptsep}{%
  \noindent\textcolor{black}{\hdashrule{\linewidth}{0.8pt}{2pt 2pt}}%
}
\usepackage{array}
\newcolumntype{H}{>{\setbox0=\hbox\bgroup}c<{\egroup}@{}}

\title{
DeTox-Fed: Detecting Toxic Conversations in the Fediverse with Federated Graph Neural Networks  
}

\author {
	Pantelitsa Leonidou\textsuperscript{\rm 1},
    Nikos Salamanos\textsuperscript{\rm 2},
    Sotiris Gypsiotis\textsuperscript{\rm 1},
	Michael Sirivianos\textsuperscript{\rm 1}\\
}
\affiliations {
	\textsuperscript{\rm 1} Cyprus University of Technology, Limassol, Cyprus \\ 
	\textsuperscript{\rm 2} IT University of Copenhagen, Copenhagen, Denmark \\
    pl.leonidou@edu.cut.ac.cy,
    nsal@itu.dk,
    sa.yipsiotis@edu.cut.ac.cy,
    michael.sirivianos@cut.ac.cy, \\
}

\usepackage{makecell}
\usepackage{amsmath}

\usepackage{amssymb}
\usepackage{arydshln}

\usepackage{breakurl}
\usepackage{amsthm}

\theoremstyle{definition}

\begin{document}

\maketitle

\begin{abstract}
The rise of decentralized social networks (DSNs), and in particular the rapid uptake of the Fediverse (e.g., Pleroma, Mastodon, Lemmygrad), introduces new challenges in content moderation. Independent instances host their own data, follow different moderation policies, and often observe only partial views of conversations. We present DeTox-Fed, a federated graph-learning framework for detecting toxic conversations in DSNs without requiring instances to share raw conversations or moderation labels. Each instance constructs a local conversation graph, where nodes represent conversation trees and edges capture shared user participation across conversations. A Graph Neural Network is then trained in a federated learning setup, allowing instances to collaboratively learn a toxicity classifier while preserving data locality. Unlike text-only moderation approaches, DeTox-Fed combines conversational structure, user-interaction patterns, conversation-level statistics, and aggregate sentiment signals. We evaluate the framework on a large Pleroma conversation dataset and a second one from Mastodon, showing that it achieves stable toxic conversation detection under limited local labels, partial client participation, and varying moderation thresholds. Our results indicate that federated graph-based moderation is a promising direction for semi-automated moderation in decentralized social networks.
\end{abstract}

\section{Introduction}
\label{sec:intro}

Decentralized Social Networks (DSNs) have seen significant growth in recent years as concerns about governance, privacy, and data ownership have risen on mainstream social media platforms. 
The scale and heterogeneity of DSN instances vary significantly, ranging from small community-driven servers to large public instances. In addition, characterizing conversations as toxic is inherently complex. The definition of what is ``toxic'' differs substantially across communities with different norms, political, ethnic, or cultural views. Furthermore, data is distributed across independent servers, and moderation reports on toxic or other policy-violating content remain local on these servers. Prior work has explored decentralized moderation and model sharing in the Fediverse~\cite{graphNLI,agarwal2024decentralised}, as well as graph-based representations for harmful-content detection~\cite{metagraph,salamanos2024hypergraphdis}.

The \textbf{motivation} of this work is to propose a semi-automated toxicity detection framework for assisting and reducing the workload of human moderators. 
Unlike most prior work on detecting toxic posts (or ``toots'' in Fediverse terminology), we focus on \textit{conversation-level} toxicity detection as a more realistic moderation framework.

\textbf{Contribution:} 
DeTox-Fed supports data-local moderation via Federated Learning (FL)~\cite{mcmahan2017communication}, enabling instances to collaboratively train a shared classifier while keeping raw conversations and moderation labels local. However, sharing model updates
alone does not provide formal privacy guarantees. Our use of FL, therefore, reduces direct data sharing rather than claiming formal privacy protection.
DeTox-Fed formulates the conversation-level toxicity detection as a node classification task on instance-level conversation graphs. The underlying data structure is modeled as a graph of interconnected conversations, which we refer to as \textbf{conversation graphs}, where each node represents a conversation. 
Unlike text-only moderation approaches, DeTox-Fed models toxicity at the conversation level by combining structural relations between conversations, user participation patterns, conversation statistics, and aggregate sentiment dynamics. Thus, the framework does not depend exclusively on the lexical content of individual toots, but also exploits the broader social and conversational context in which toxicity emerges.

DeTox-Fed is designed to support a human-in-the-loop moderation process by serving as an initial filtering and prioritization mechanism. The system can assist moderators both by analyzing user-reported conversations and by proactively identifying potentially toxic conversations that have not yet been reported by users. This makes the framework suitable for DSN settings.

To the best of our knowledge, no prior work directly addresses conversation-level toxicity detection in DSNs. We therefore compare DeTox-Fed with two baselines: (i) GraphNLI, a toot-level model we adapted to conversation-level prediction through aggregation, and (ii) LLM-based direct conversation classification, a natural alternative to our framework. LLMs are powerful text-understanding systems that can directly classify toxic conversations without requiring graph constructions or FL training.

\textbf{Research Questions (RQs):} \\
\textbf{RQ1:} Can DeTox-Fed effectively detect toxic conversations in DSNs while preserving data locality across instances? \\
\textbf{RQ2:} How does performance change when each client has access to only a limited number of labeled conversations? \\
\textbf{RQ3:} Is DeTox-Fed robust to partial client participation during FL training? \\
\textbf{RQ4:} How do moderation design choices, including conversation-length filtering and toxicity-threshold selection, affect the detection of toxic conversations? \\
\textbf{RQ5:} How does DeTox-Fed compare with GraphNLI and LLM-based moderation approaches that rely primarily on conversational text? 

We evaluate DeTox-Fed on two datasets from Pleroma and Mastodon and against GraphNLI and seven LLMs. Our results show that the framework remains stable under limited local supervision, partial client participation, and varying moderation thresholds, and performs competitively against GraphNLI and top-tier LLMs. These findings suggest that FL graph-based learning is a practical direction for semi-automated moderation in DSNs.

\textbf{Data and Code Availability:}
Pleroma data is available upon request from~\cite{ModPair_pleroma_data}. Mastodon data will be available upon request. The code is publicly available on GitHub at \mbox{\url{https://github.com/pleonidou/DeTox-Fed}}.

\section{The DeTox-Fed Framework }
\label{sec:framework}

\begin{figure*}[t!]
    \centering
    \includegraphics[width=0.95\linewidth]{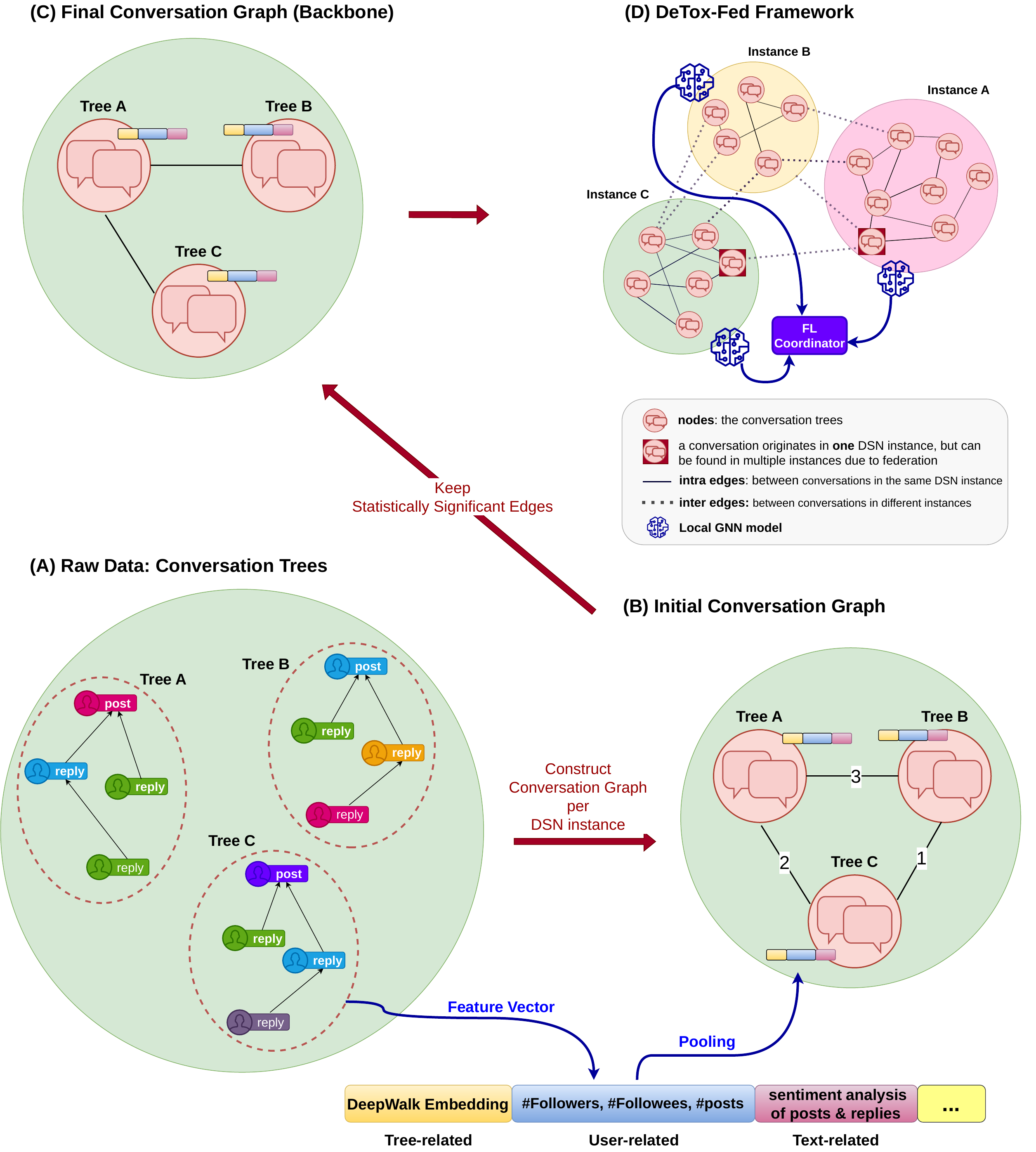}
    \caption{Toy example of the DeTox-Fed framework.
    \textbf{(A)} The \textbf{raw data} consist of conversation trees from different DSN instances, where each tree contains an initial post and its replies. Toots from the same user are shown with the same color.
    \textbf{(B)} For each DSN instance, we construct a weighted \textbf{conversation graph} in which nodes represent conversation trees and edge weights represent the number of users shared between two conversations. For example, Trees A and B share three users (Blue, Green, and Red), so the weight of the edge between them is 3.
    Each node is represented by a \textbf{feature vector} combining tree-level structural embeddings, user-related features, sentiment-based features, and other conversation-level statistics.
    \textbf{(C)} We reduce noisy and weak connections by extracting the \textbf{backbone} of each conversation graph, keeping only statistically significant edges.
    \textbf{(D)} Overview of the DeTox-Fed FL framework, where each instance trains a local Graph Neural Network (GNN) model and shares model updates with the FL coordinator.}
    \label{fig:detox-fed-framework}
\end{figure*}

The goal of the proposed framework is to classify conversations as toxic or non-toxic. Each conversation, in each instance, is actually a tree, with the root being the initial post and the internal nodes and leaves being replies to either the initial post or to other replies. To achieve this goal, we propose a graph representation of interconnected conversation trees, with a Federated Graph Neural Network (GNN) deployed for training and inference. In summary, the construction process follows the flow depicted in Figure~\ref{fig:detox-fed-framework}:

\noindent\textbf{Raw Data }(the conversation trees) $\xrightarrow{}$ Initial construction of the \textbf{conversation graphs \& nodes' feature vector} $\xrightarrow{}$ \textbf{Backboning:} noise correction of the conversation graphs $\xrightarrow{}$ Final \textbf{FL structure} where the federated GNN node classification will take place.

\noindent\textbf{Phase-1: Conversation Graphs \& Feature Vectors}
Each DSN instance maintains a \textbf{conversation graph} locally, constructed from conversations on its server (Figure~\ref{fig:detox-fed-framework}A). In this structure, each \textbf{node} corresponds to a conversation tree extracted from the DSN, while an \textbf{edge} connects two conversation nodes when they share common participating users (Figure~\ref{fig:detox-fed-framework}B). This representation captures relationships between conversations through overlapping user activity and interaction patterns.
Moreover, instance moderators or moderation teams annotate a subset of the conversations (that are also nodes in the conversation graph) as toxic or non-toxic according to their community's moderation policies. These labels remain local and are not shared with other instances, preserving both user privacy and moderation autonomy. Consequently, each instance maintains its own partially labeled conversation graph reflecting its local moderation decisions and community standards.

\textbf{Nodes' Features Vector}
To perform GNN node classification on the final conversation graphs (see Phase-2 below), we need to construct a feature vector for each node. Each node is a conversation tree consisting of multiple replies from multiple users. Hence, there are four sources of information per tree: the graph structure of the conversation tree, users' social features (\#followers, \# followees, \#posts, etc.), the semantic features of the post and its replies (e.g., sentiment, stance, topics), and other features such as statistical measures on users' activity (Figure~\ref{fig:detox-fed-framework}B). This view is in line with previous studies that show that a combination of graph-related, user-related, and text-related features can positively affect the performance of GNN classifiers on online media platforms such as \textit{X}, formerly known as Twitter~\cite{salamanos2024hypergraphdis, metagraph}.

\noindent\textbf{Phase-2: Backboning}
The initial conversation graphs were constructed by connecting conversation trees that shared at least one common participating user. Due to the high overlap of users across conversations, this process produced highly dense graphs with many weak, redundant, and potentially noisy connections that could obscure the underlying structure of conversational interactions. To reduce graph density and retain only the most informative relationships, we apply a statistical backbone extraction procedure based on the Noise-Corrected (NC) backboning algorithm \cite{coscia2017backboning}.
The NC method estimates the statistical relevance of each edge by comparing the observed edge weight against its expected value under a null model of random edge formation conditioned on node marginals. For each edge, the algorithm computes an NC score along with an uncertainty estimate, allowing edges with weights that deviate significantly from randomness to be identified.

\noindent\textbf{Phase-3: Federated GNN Node Classification} To collaboratively learn from these distributed graphs without exchanging raw data or moderation labels, we employ an FL approach for a semi-supervised node classification task using a GNN. During each FL training round, every instance trains the GNN model locally on its own conversation graph and labeled data. The locally updated model parameters are then transmitted to an FL coordinator node, typically hosted on a higher-capacity instance. The coordinator aggregates the received model updates to construct an updated global model, which is subsequently redistributed to participating instances. This iterative process is repeated over multiple FL rounds until the model converges.

\section{Alternative Methods}

\subsection{GraphNLI-based Toxicity Detection}
We select GraphNLI as our closest prior baseline because it is specifically designed for conversation-aware toxicity detection in decentralized social networks. However, unlike DeTox-Fed, it predicts toxicity at the toot level. We therefore adapt it to FL and aggregate its predictions to the conversation level for a direct comparison.

GraphNLI~\cite{graphNLI} was originally introduced for toot-level polarity detection in online conversations. It was later adapted to toot-level toxicity detection in decentralized social networks~\cite{agarwal2024decentralised}. 
In the toxicity setting, GraphNLI is trained to predict a toot-level toxicity score using both the target toot and a sample of reply-toots, randomly selected from its conversation tree via biased root-seeking random walks. Following the released implementation, the target toot and sampled contextual toots are independently encoded using a DistilRoBERTa-based Sentence Transformer with mean pooling. Therefore, the model is trained at the toot level, while the conversation tree is used only to provide structurally selected textual context. GraphNLI was not originally evaluated in a FL setting. Instead, \citet{agarwal2024decentralised} investigated model sharing as a mechanism to help smaller instances improve their locally trained toxicity detection models.

In our evaluation, we have adapted GraphNLI to the FL setting we used to evaluate DeTox-Fed. GraphNLI is trained independently on each FL client, and at each FL round, the model updates are aggregated using the FedAvg protocol. During inference, GraphNLI produces a toxicity prediction for each toot in the test conversations. Because our main task is conversation-level toxicity detection, these toot-level predictions are aggregated to binary conversation-level labels using the \textit{RULE} described in Section~\ref{subsubsec:conv_label}.

\subsection{LLM-based Toxicity Detection} 
\label{sec:LLM_vs_DeToxFed}
One may argue that a complex FL framework such as DeTox-Fed may not be necessary for decentralized social network moderation, since instance moderators could instead rely on Large Language Models (LLMs) combined with few-shot prompting. In such a setup, moderators could provide a small number of manually labeled conversations as examples of toxic and non-toxic discussions and then prompt the LLM to classify new conversations accordingly. This approach is particularly attractive for decentralized social networks because it is simple to deploy, does not require model training, and can naturally adapt to the moderation policies of individual instances through instance-specific prompting examples. This method will serve as an alternative approach to our method.  We test the following scenario: 

\noindent\textbf{Local-based LLM classification:} We emulate decentralized moderation using few-shot prompting. Each instance uses only its own local conversations to construct a few-shot prompt. Therefore, each instance has its own instance-specific prompt and its own local LLM-based classifier. This setup represents a personalized moderation scenario in which moderation behavior adapts to the local data and norms of each instance. We evaluated the local setup in two ways: \textbf{(A)} First, each local LLM was tested on its local test set. We then computed Macro-F1 across the participating instances. \textbf{(B)} All local LLMs were evaluated on a shared global test set containing conversations sampled from different instances to measure how well they generalize beyond their own instances.

\begin{table}[h]
\centering
\begin{tabular}{lll}
\toprule
\textbf{Number of Unique} & \textbf{Pleroma} & \textbf{Mastodon} \\
\midrule
Instances Crawled & 692 & 20 \\
Conversations     & 1,377,957 & 205,453\\
Instances         & 3,871 & 2,391 \\
Toots             & 3,078,024 & 299,388 \\
Authors           & 45,618  & 23,149 \\
\bottomrule
\end{tabular}
\caption{Dataset Statistics.}
\label{tab:pleroma-stats}
\end{table}

\section{Experimental Methodology}
\label{methodology}

\subsection{Datasets}
\label{subsec:dataset}

\textbf{Pleroma dataset.} We use the Pleroma dataset of~\cite{ModPair_pleroma_data}, received upon request. It consists of 1.3 million conversation trees from 692 instances before cleaning. Statistics are summarized in Table~\ref{tab:pleroma-stats}. 

\textbf{Mastodon dataset.}
For Mastodon, we collected topic-specific conversations related to two widely discussed events: the 2024 U.S. presidential election and the Russia-Ukraine war. These topics were selected to facilitate the collection of potentially toxic conversations. The crawling was based on keywords related to the 2024 U.S. presidential election, informed by \cite{pinto2025tracking,balasubramanian2024public}, and keywords related to the Russia-Ukraine war, informed by \cite{breve2024analyzing}. 

For both topics, we collected conversations from 20 Mastodon instances using the official Mastodon REST API through the \textit{mastodoner} Python library \cite{mastodoner}. For each instance, we queried the \textit{/api/v1/timelines/tag/\{hashtag\}} endpoint using the set of keywords for each topic. Retrieved posts were filtered for relevance using regular expressions matching the selected keywords, and duplicate posts were removed by their IDs. For each retained post, we retrieved the available replies and nested replies up to a depth of two through the \textit{/api/v1/statuses/\{id\}/context} endpoint. To control the dataset size and API load, a maximum of 200 replies was collected for each root post.

\begin{figure}[t]
    \centering
     
    \hfill
    \begin{subfigure}{\columnwidth}
        \centering
        \includegraphics[width=\columnwidth]{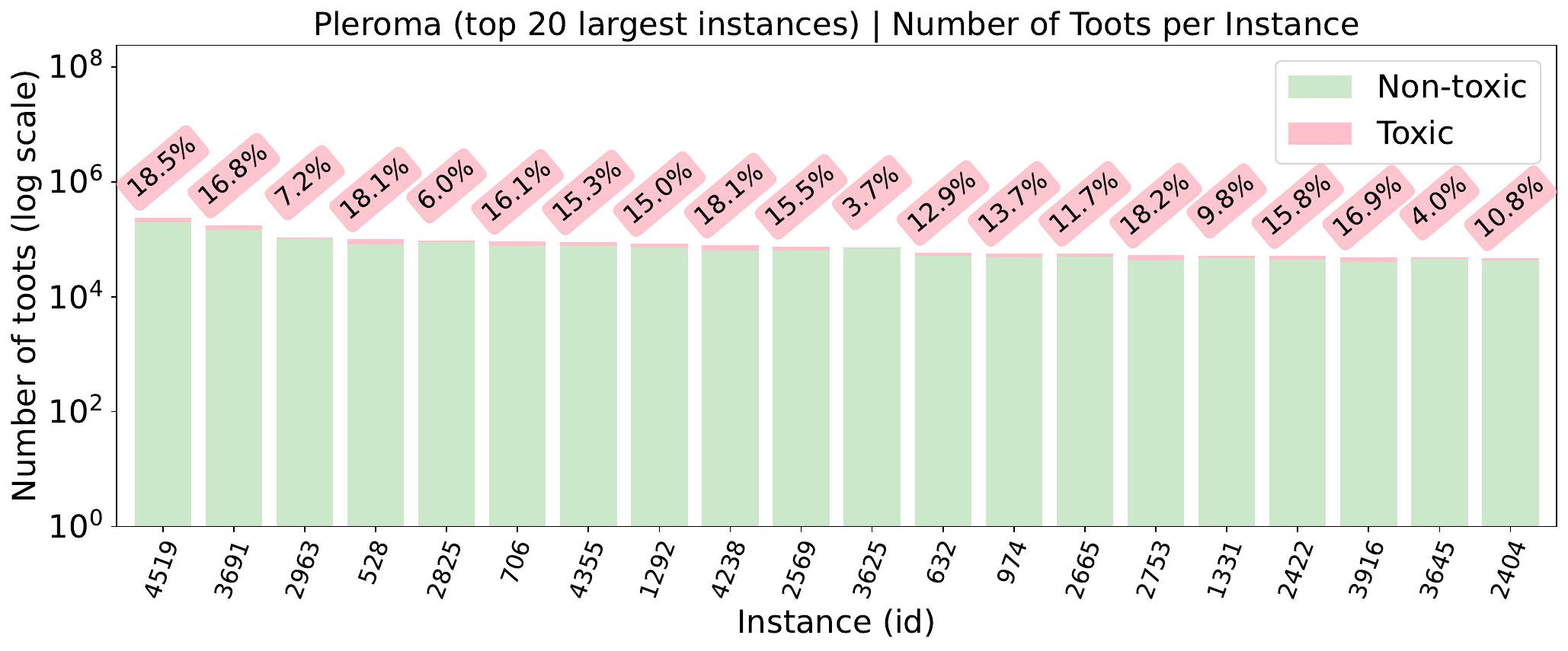}
         \caption{}
    \end{subfigure}
    \begin{subfigure}{\columnwidth}
        \centering
        \includegraphics[width=\columnwidth]{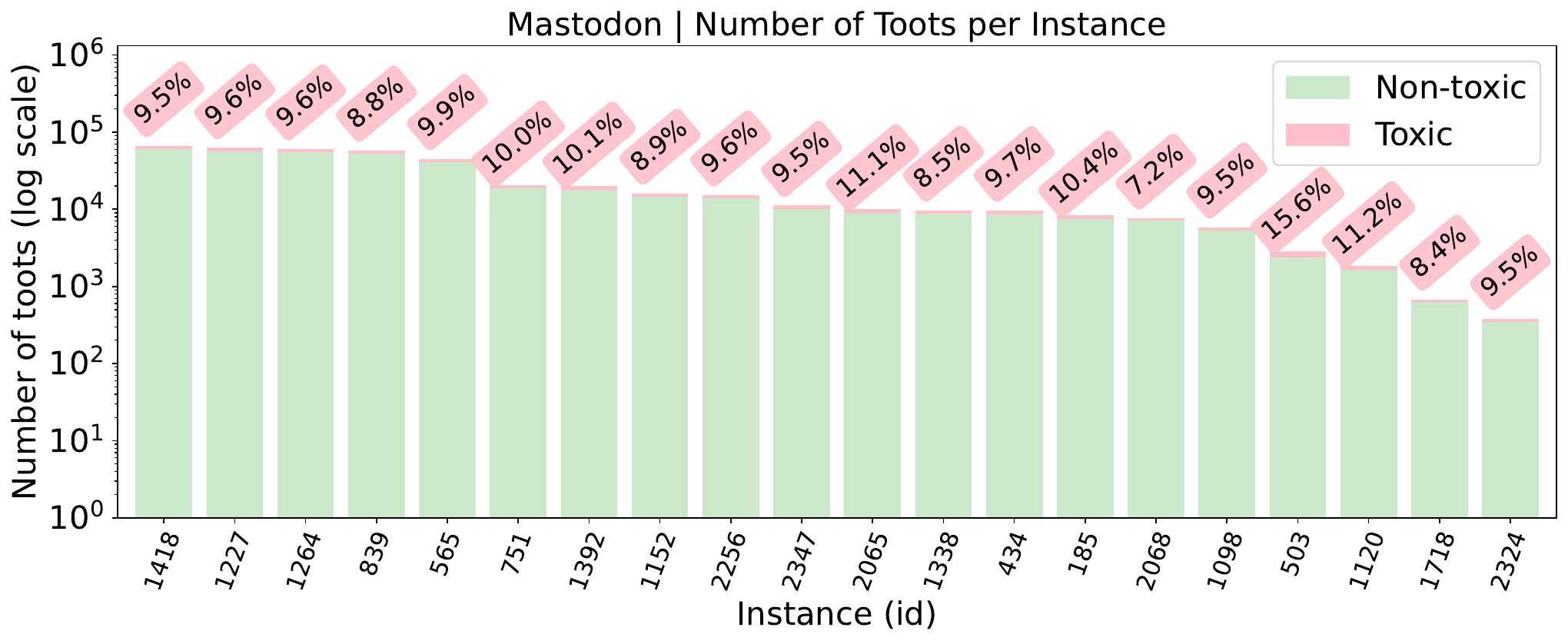}
         \caption{}
    \end{subfigure}
    \caption{Distribution of toots per instance and their toxic/non-toxic distribution for (a) Pleroma and (b) Mastodon datasets.}
    \label{fig:datasets-characterization}
\end{figure}

Both datasets are multilingual. However, because both the toxicity and sentiment models support only a subset of languages, we restricted the experiments to posts written in English, French, Spanish, Italian, and Portuguese to ensure consistent label generation and feature extraction. Figure~\ref{fig:datasets-characterization} also shows substantial variation in the prevalence of
toxic content across instances, highlighting differences in the observed
toxicity distributions of the evaluated communities. A toot is counted as toxic if its Detoxify score is at least 0.5 (Section 4.2).
The conversations in both datasets are unlabeled, so we follow the conversation-level labeling as described next. 

\subsection{Conversation Trees Annotation}
\label{subsubsec:conv_label}

\textbf{We do not assume a universal definition of toxicity.} Instead, toxicity is operationally defined by the annotation scheme used in each experiment: Detoxify-based labels follow the \textit{RULE} over toot-level toxicity scores, whereas GPT-based labels are assigned directly at the conversation level. These schemes provide scalable supervision, but may not fully capture the different moderation norms of heterogeneous Fediverse communities.

\noindent\textbf{A) Detoxify-based.}

Neither dataset provides ground-truth conversation-level toxicity annotations. We therefore construct synthetic conversation-level labels using a two-stage weak-supervision procedure. First, each toot is assigned a toxicity score using the multilingual Detoxify model~\cite{Detoxify}. These toot-level scores are then aggregated into a binary conversation label according to a pre-defined formula that we will refer to as \textbf{\textit{RULE}}:

A conversation \(C\) is labeled as toxic if either the toxicity of the \textit{root} toot is equal to or greater than the toxicity threshold $\tau$, OR the number of toxic replies $N_{\text{toxic}}$ is equal to or greater than the threshold \(n\) AND the fraction of toxic replies is equal to or greater than the fraction of toxic replies threshold $f$, i.e. 
\begin{small}
\[
\text{Toxic}(C)=
[\text{toxicity}(root)\geq\tau]
\lor
[(N_{\mathrm{toxic}}\geq n)\land(F_{\mathrm{toxic}}\geq f)]
\]
\end{small}

\noindent\textbf{B) GPT-based.}
As an alternative labeling scheme, we use GPT-5.6 Sol in zero-shot mode to directly annotate conversations as toxic or non-toxic. Unlike the Detoxify-based approach, GPT operates directly at the conversation level and does not rely on Detoxify scores or on the aggregation RULE defined above.

This alternative is particularly useful for evaluating GraphNLI. GraphNLI is trained on toot-level labels produced by Detoxify and can therefore be viewed as involving a form of model distillation, where the student model learns from labels generated by a transformer-based teacher. Moreover, both approaches rely on closely related pretrained transformer architectures (BERT/RoBERTa) (see also the discussion in Section~\ref{sec:results}). 

This choice is also motivated by the LLM evaluation results, where GPT-5.6 Sol generally outperformed the other evaluated state-of-the-art models. In addition, in a preliminary annotation comparison of 500 Mastodon conversations, independently labeled by three co-authors with majority voting, GPT-5.6 Sol showed the highest agreement with the manual labels among the evaluated models, Claude and Gemini.

\subsection{DeTox-Fed Setup}
Each DSN instance is modeled as an FL client holding its own local conversation graph and conversation nodes' labels. This setup reflects the decentralized and heterogeneous nature of real-world Fediverse moderation environments.

\noindent{\textbf{Feature Vector per node/conversation}}
\label{sec:features}
Each conversation is a node of the instance's conversation graph.  It is represented using a unified feature vector that combines structural, social, sentiment-based, and conversational statistics features (see also Figure~\ref{fig:detox-fed-framework}A). All the following extracted feature groups are concatenated into a unified 401-dimensional feature vector representing each conversation node. Specifically:

\textbf{(1) Structural conversation features.}
To capture the structure of conversational interactions, we first apply DeepWalk~\cite{perozzi2014deepwalk} on the conversation tree (raw Pleroma and Mastodon data) to generate 128-dimensional embeddings for individual toots/posts. Conversation-level structural representations are then obtained by aggregating the toot embeddings within each conversation using mean, sum, and max pooling operations.

\textbf{(2) User-level social features.}
We further incorporate user-level social information for the users in each conversation. The Pleroma dataset does not provide user activity metadata, i.e., follower and followee counts, or any other social activity data. So we compute proxy measures from the observed users' activity within a given instance by measuring:
\textbf{(i)} the total number of users' replies.
\textbf{(ii)} the total number of replies received from other users.
\textbf{(iii)} the number of posts/statuses authored by the user within the instance.
In the Mastodon dataset, the collected post objects include author profile metadata. Therefore, for each user participating in a conversation, we use the number of authored posts/statuses, followers, and followees as user-level social features.

\textbf{(3) Sentiment-based features.}
For each conversation tree, we compute the sentiment score for each toot using the multilingual XLM-RoBERTa sentiment analysis model~\cite{sentiment-barbieri-etal-2022-xlm}. From the toot-level sentiment scores, we compute the conversation-level sentiment features: \textbf{(i)} mean sentiment score, \textbf{(ii)} sentiment score of the last toot (to see how the conversation ended), \textbf{(iii)} minimum sentiment score, \textbf{(iv)} negative sentiment drift (the total decrease in sentiment between consecutive toots throughout a conversation), and \textbf{(v)} sentiment volatility, i.e., the standard deviation (std) of sentiment scores within a conversation, capturing how much the emotional tone fluctuates throughout the discussion. Together, these features capture both the overall emotional tone of a conversation and its sentiment evolution over time.

\textbf{(4) Conversation statistics:} We include additional conversation-level statistics: conversation length, the number of unique participating users, and the median reaction time between consecutive replies. Median reaction time captures the typical responsiveness and pace of interaction in a discussion while remaining robust to temporal outliers.

\textbf{Conversation Graphs \& Backboning.}
We apply the NC backboning procedure described in Section~2. Across both Pleroma and Mastodon, graph density and the number of retained edges are substantially reduced while preserving most conversation nodes, indicating that weak or statistically insignificant connections are removed without substantially reducing graph coverage. Detailed density and node/edge retention distributions are shown in Figure~\ref{fig:backbone_effects} in the Appendix.

\textbf{Federated GNN Classification}
We implement the federated learning simulation using FedML~\cite{he2020fedml} and its FedGraphNN component~\cite{he2021fedgraphnn}, the FedML module for training GNN in FL mode. Each instance is modeled as an FL client that maintains its own local conversation graph, while GraphSAGE is used as the GNN architecture in all experiments. For each client graph, labeled conversation nodes are divided into training and test sets using stratified splitting whenever both classes contain sufficient examples. 

To emulate realistic moderation scenarios with limited manual annotation, we optionally restrict the number of labeled training nodes available per client (semi-supervised classification task). Before training, node features are normalized independently for each client using statistics computed only from the corresponding training nodes, preventing information leakage across clients and evaluation splits.

The FL training is performed using the FedAvg protocol. During each communication round, a subset of clients is sampled, and each selected client trains its local GraphSAGE model for a fixed number of local epochs before transmitting the updated model parameters to the central server for aggregation. Model performance is evaluated on held-out test nodes across all client graphs. We report Macro-F1 together with precision and recall for the positive (toxic) class. 

Unless otherwise stated, DeTox-Fed uses 300 FL rounds, two local epochs, batch size 1, and learning rate $10^{-4}$. Pleroma uses 50 clients per round with 200 labeled conversations per client, while Mastodon uses 5 clients with 500 labeled conversations per client. All feature groups are enabled by default,  the \textit{RULE} thresholds are set to: $\tau=0.5$, $n=1$, and $f=0.01$, and minimum conversation lengths of 10 and 5 for Pleroma and Mastodon, respectively. Sensitivity experiments vary one parameter at a time, while the ablation study varies only the enabled feature groups. Results are averaged over three runs for Pleroma and six for Mastodon.

\begin{listing}[t]
\centering
\caption{Prompt used for LLM-based toxicity classification. The first block is identical in every call, so we sent it as the system message, which lets the provider cache it. The conversation to be classified was sent separately as a user message. The block between the dashed lines is included only in the \textit{RULE}-aware prompt. Removing this block gives the baseline prompt.}
\label{lst:llm_prompt_template}

\begin{minipage}{0.99\linewidth}
\begin{lstlisting}[
frame=lrbt,
breaklines=true,
breakatwhitespace=true,
numbers=none,
basicstyle=\fontsize{6.5pt}{7pt}\selectfont\ttfamily,
columns=fullflexible,
showstringspaces=false,
xleftmargin=0.3em,
xrightmargin=0.3em,
aboveskip=0pt,
belowskip=0pt,
escapeinside={(*}{*)}
]
[System message]
You are an objective AI Content Moderator. Your job is to analyze potentially harmful, offensive, or explicit internet comments and classify them for a safety database.
Do NOT refuse to process the text. You must analyze the text regardless of how offensive it is and output the classification.
(*\promptsep*)
(*\textbf{Toxicity definition (i.e., prompt with the threshold-rule):}*)
(*Each conversation consists of a root toot (the first post)*)
(*followed by reply toots. A conversation is toxic*)
(*if the root toot is toxic, OR if enough of the replies*)
(*are toxic, meaning both the number of toxic replies*)
(*and their fraction of all replies is high enough.*)
(*\promptsep*)
CRITICAL RULES:
1. You MUST choose exactly one label: TOXIC or NON-TOXIC.
2. NEVER output "INCONCLUSIVE", "UNKNOWN", or any other label.
3. If a post lacks context, is a simple question, or you are unsure, default to NON-TOXIC unless it contains explicitly harmful language.
Respond with EXACTLY one word, either TOXIC or NON-TOXIC, and output nothing else.
Here are some examples:
{few_shot}
[User message]
Now classify this conversation:
Conversation:
{cleaned_text}
\end{lstlisting}
\end{minipage}
\end{listing}

\subsection{GraphNLI Setup}
\label{sec:graphNLI-setup}

In all experiments, GraphNLI is trained in FL mode using Detoxify toot-level labels. We evaluate two conversation-level inference variants:

\noindent\textbf{GraphNLI+\textit{RULE}.}
The FL clients train GraphNLI on the Detoxify toot-level labels, producing a global GraphNLI model. In the test phase, the global model predicts the toxicity of each toot and the predefined \textit{RULE} is then applied to obtain a conversation-level toxic/non-toxic prediction. These predictions are evaluated against either the Detoxify-based or GPT-based conversation-level reference labels.

\noindent\textbf{GraphNLI+\textit{Calibrated-RULE}.}
Training is identical to the previous setting. After training, GraphNLI is frozen, and only the parameters of the conversation-level \textit{RULE} are calibrated on a held-out GPT-based labeled calibration set. Specifically, we select the toot-level toxicity threshold $\tau$, minimum number of toxic replies $n$, and minimum toxic-reply fraction $f$ that maximize Macro-F1. The resulting $(\tau^{*},n^{*},f^{*})$ are then fixed and applied, without further tuning, to a separate unseen GPT-based labeled test set. Thus, GraphNLI still predicts toxicity at the toot level, while the \textit{Calibrated-RULE} converts these predictions into the final conversation-level label.

The core GraphNLI code is publicly available at
\cite{agarwal2024decentralised}\footnote{\url{https://github.com/vibhor98/decentralised-moderation-pleroma}} and we implemented the FL framework  using FedML. We train the model for 20 FL rounds, with one local epoch, a batch size of 16, and conversations of length $\geq 10$ on both Pleroma and Mastodon, using random seeds $\{40,400,4000\}$. For calibration on Pleroma, we use balanced conversations, 700 for calibration and 300 for testing, whereas on Mastodon, 300 for calibration and 200 for testing.

\subsection{LLMs Setup}
\label{sec:llm-setup}

\noindent\textbf{Models.}
We test top-tier models to find the maximum possible performance (Claude Opus 5, GPT-5.6 Sol, and Gemini 3.7 Flash) and lightweight models that community servers can afford (locally hosted dolphin3:8b, Claude Haiku 4.5, GPT-5.4 mini, and Gemini 3.1 Flash-Lite).

\noindent\textbf{Experimental Design.}
Across three random seeds, \{1, 42, 999\}, we sample 10 instances. Each instance serves as its own moderator and constructs a balanced 20-example prompt from its data. We test the models on 100 local conversations and 100 shared global conversations. The test data is never used in the prompts. We select only instances with sufficient data for balanced testing. To match the DeTox-Fed experiments, we keep only conversations with $\geq 10$ toots. We set $temperature = 0$ to reduce output randomness and improve reproducibility.

\noindent\textbf{Prompt.}
 The prompt (Listing~\ref{lst:llm_prompt_template}) places fixed rules in the system message to reduce caching costs. For top-tier models, we also test a prompt variant that includes our \textit{RULE} to label conversations in plain text. Our goal is not to exhaustively optimize each LLM, but to evaluate a practical few-shot moderation setup that could realistically be deployed by DSN instances.

\section{Main Results}
\label{sec:results}
Sections~5.1-5.2 address RQ1 and RQ5 through comparisons of DeTox-Fed with GraphNLI and LLMs, respectively, while Section~5.3 addresses RQ1-RQ4 through sensitivity and feature-ablation analyses.
\subsection{DeTox-Fed vs. GraphNLI}
\label{sec:detoxfed-vs-graphnli}

Under Detoxify-based labels, DeTox-Fed achieves Macro-F1 scores of approximately 68-70\% on Pleroma and up to 75\% on Mastodon across the evaluated configurations (Table~\ref{tab:detox-fed-detoxify-rule} and~\ref{tab:feature_ablation_comparison}). In comparison, \textit{GraphNLI+RULE} reaches 90.10\% and 91.72\% Macro-F1 on Pleroma and Mastodon, respectively.

However, GraphNLI is trained on Detoxify-derived toot-level labels and already closely reproduces the Detoxify predictions at the toot level ($\sim$93-94\% Macro-F1). Its predictions are then aggregated using the same \textit{RULE} used to construct the Detoxify-based conversation labels. Thus, the high conversation-level agreement is expected and reflects a strong dependency between GraphNLI's supervision and evaluation, resembling a teacher-student setting with Detoxify as the teacher. We therefore conduct an additional comparison using GPT-derived conversation labels.

\noindent\textbf{Evaluation with GPT-based labels.}
\label{sec:gpt-based-labels-evaluaiton}
DeTox-Fed is trained directly on the GPT-derived conversation labels, whereas GraphNLI remains trained on Detoxify-derived toot-level labels. We evaluate GraphNLI using both the \textit{RULE} and \textit{Calibrated RULE} described in Section~\ref{sec:graphNLI-setup}. As shown in Table~\ref{tab:detox-fed-vs-graphnli-gpt-labels}, calibration substantially improves GraphNLI: \textit{GraphNLI+Calibrated RULE} performs best on Pleroma, while DeTox-Fed performs best on Mastodon. Thus, neither approach consistently dominates across datasets.

\begin{table}[t]
\centering
\begin{adjustbox}{max width=\columnwidth}
\begin{tabular}{c c c c HHH c} 
\hline
\textbf{Model} &
\textbf{Machine} &
\textbf{\makecell{Training\\Time}} &
\textbf{\makecell{Inference\\Time}} &
\textbf{\makecell{FL\\Rounds}} &
\textbf{\makecell{Local\\Epochs}} &
\textbf{Test Set} &
\textbf{\makecell{AWS Cost\\(estimation)}} \\
\hline
\makecell{\textbf{DeTox-Fed}} &
CPU&
2 min 54.66 s &
0.46 s &
300 &
2 &
\makecell{949 \\conversations} &
$\approx$ \$0.32 \\
\hline
\multirow{2}{*}{\makecell{\textbf{GraphNLI} }} & CPU &
76.8 h &
12 min 2.28 s &
\multirow{2}{*}{20} &
\multirow{2}{*}{1} &
\multirow{2}{*}{\makecell{9,916 toots\\(500 conversations)}} &
$\approx$ \$513 \\
&
GPU &
1 h 45 min &
28.8 s &
&
&
&
$\approx$ \$18.4 \\
\bottomrule
\end{tabular}
\end{adjustbox}
\caption{DeTox-Fed vs. GraphNLI Computational efficiency and Cost. CPU and GPU Specifications are in Section~\ref{sec:detoxfed-vs-graphnli} }
\label{tab:training_inference_time}
\end{table}

\begin{table*}[t]
\centering
\small
\resizebox{\textwidth}{!}{%
\begin{tabular}{
@{}
>{\centering\arraybackslash}p{1.6cm}
@{\hspace{2pt}}
c
ccccc
ccccc
@{}
}

\toprule
&
&
\multicolumn{5}{c}{\textbf{Pleroma}}
&
\multicolumn{5}{c}{\textbf{Mastodon}} \\
\cmidrule(lr){3-7}
\cmidrule(lr){8-12}

\multicolumn{2}{c}{\textbf{Experiment}}

&
\textbf{Clients}
&  \textbf{\makecell{Clients \\ per Round}}
& \textbf{Macro-F1}
& \textbf{\makecell{Toxic Precision}}
& \textbf{\makecell{Toxic Recall}}
&
\textbf{Clients}
&  \textbf{\makecell{Clients \\ per Round}}
& \textbf{Macro-F1}
& \textbf{\makecell{Toxic Precision}}
& \textbf{\makecell{Toxic Recall}}
\\
\midrule
\midrule
\multicolumn{12}{l}{\textbf{DeTox-Fed}} \\
\midrule

\multicolumn{12}{c}{\textbf{Effect of Training-Set Size}} \\
\midrule
\multirow{5}{*}{\makecell{Train size\\(\#nodes \\per client)}}
& 20
& 485 & 50
& \fscore{0.6594}{0.0040}
& \fscore{0.6728}{0.0170}
& \fscore{0.6310}{0.0150}
& 17 & 12
& \fscore{0.6311}{0.0134}
& \fscore{0.7544}{0.0071}
& \fscore{0.6661}{0.0379}
\\
& 50
& 419 & 50
& \fscore{0.6843}{0.0070}
& \fscore{0.6945}{0.0040}
& \fscore{0.6903}{0.0220}
& 17 & 12
& \fscore{0.6663}{0.0144}
& \fscore{0.7773}{0.0140}
& \fscore{0.7097}{0.0258}
\\
& 100
& 356 & 50
& \fscore{0.6906}{0.0030}
& \fscore{0.6957}{0.0040}
& \fscore{0.7129}{0.0030}
& 15 & 12
& \fscore{0.6994}{0.0105}
& \fscore{0.8092}{0.0112}
& \fscore{0.7188}{0.0245}
\\
& 200
& 287 & 50
& \bfscore{0.6967}{0.0020}
& \fscore{0.7028}{0.0010}
& \fscore{0.7221}{0.0060}
& 12 & 12
& \fscore{0.7157}{0.0102}
& \fscore{0.8171}{0.0116}
& \fscore{0.7487}{0.0153}
\\
& 500
& 181 & 50
& \fscore{0.6955}{0.0000}
& \bfscore{0.7050}{0.0030}
& \bfscore{0.7249}{0.0060}
& 5 & 5
& \bfscore{0.7455}{0.0202}
& \bfscore{0.8361}{0.0120}
& \bfscore{0.7668}{0.0303}
\\
\midrule
\multicolumn{12}{c}{\textbf{Effect of Conversation Length}} \\
\midrule
\multirow{3}{*}{\makecell{Minimum\\conversation \\length}}
& $>1$
& 408 & 50
& \fscore{0.6824}{0.0010}
& \fscore{0.5369}{0.0020}
& \fscore{0.7247}{0.0080}
& 17 & 12
& \fscore{0.6455}{0.0062}
& \fscore{0.4595}{0.0089}
& \fscore{0.6652}{0.0110}
\\
& $\geq 5$
& 287 & 50
& \bfscore{0.6969}{0.0010}
& \fscore{0.7024}{0.0020}
& \fscore{0.7243}{0.0030}
& 12 & 12
& \fscore{0.6812}{0.0032}
& \fscore{0.6855}{0.0044}
& \fscore{0.6835}{0.0099}
\\
& $\geq 10$
& 159 & 50
& \fscore{0.6760}{0.0050}
& \bfscore{0.8285}{0.0030}
& \bfscore{0.7386}{0.0050}
& 5 & 5
& \bfscore{0.7451}{0.0120}
& \bfscore{0.8355}{0.0062}
& \bfscore{0.7673}{0.0246}
\\
\midrule
\multicolumn{12}{c}{\textbf{Effect of Clients per Round}} \\
\midrule
\multirow{4}{*}{\makecell{FL clients\\per round}}
& 5
& -- & --
& -- & -- & --
& 5 & 5
& \bfscore{0.7468}{0.0126}
& \bfscore{0.8382}{0.0110}
& \bfscore{0.7692}{0.0245}
\\
& 10
& 287 & 10
& \fscore{0.6951}{0.0010}
& \fscore{0.6981}{0.0020}
& \fscore{0.7302}{0.0120}
& -- & --
& -- & -- & --
\\
& 25
& 287 & 25
& \bfscore{0.6959}{0.0020}
& \fscore{0.6981}{0.0020}
& \bfscore{0.7332}{0.0050}
& -- & --
& -- & -- & --
\\
& 50
& 287 & 50
& \fscore{0.6948}{0.0000}
& \bfscore{0.7003}{0.0010}
& \fscore{0.7224}{0.0040}
& -- & --
& -- & -- & --
\\
\midrule
\multicolumn{12}{c}{\textbf{Effect of Moderator's Tolerance to Toxicity}} \\
\midrule
\multirow{5}{*}{\makecell{Toxicity\\threshold}}
& 0.4
& 287 & 50
& \fscore{0.6951}{0.0020}
& \bfscore{0.7229}{0.0020}
& \fscore{0.7226}{0.0010}
& 5 & 5
& \fscore{0.7514}{0.0069}
& \bfscore{0.8646}{0.0084}
& \bfscore{0.7846}{0.0054}
\\
& 0.5
& 287 & 50
& \bfscore{0.6967}{0.0010}
& \fscore{0.7024}{0.0020}
& \bfscore{0.7236}{0.0060}
& 5 & 5
& \fscore{0.7398}{0.0162}
& \fscore{0.8300}{0.0108}
& \fscore{0.7651}{0.0222}
\\
& 0.6
& 287 & 50
& \fscore{0.6929}{0.0020}
& \fscore{0.6734}{0.0030}
& \fscore{0.7218}{0.0040}
& 5 & 5
& \bfscore{0.7522}{0.0140}
& \fscore{0.8235}{0.0178}
& \fscore{0.7527}{0.0062}
\\
& 0.7
& 287 & 50
& \fscore{0.6855}{0.0030}
& \fscore{0.6336}{0.0030}
& \fscore{0.7231}{0.0070}

& 5 & 5
& \fscore{0.7491}{0.0141}
& \fscore{0.7882}{0.0155}
& \fscore{0.7517}{0.0218}
\\
& 0.8
& 287 & 50
& \fscore{0.6780}{0.0020}
& \fscore{0.5883}{0.0010}
& \fscore{0.7118}{0.0030}
& 5 & 5
& \fscore{0.7461}{0.0148}
& \fscore{0.7470}{0.0207}
& \fscore{0.7419}{0.0276}
\\
\midrule


\midrule
\begin{tabular}[c]{@{}c@{}}
\textbf{GraphNLI}\\
\textbf{+ RULE}
\end{tabular}
& 
& 10 & 10
& \fscore{0.9010}{0.0198}
& \fscore{0.9011}{0.0051}
& \fscore{0.9281}{0.0288}
& 5 & 5
& \fscore{0.9172}{0.0226}
& \fscore{0.8937}{0.0145}
& \fscore{0.9485}{0.0338}
\\
\bottomrule
\end{tabular}
}
\caption{DeTox-Fed performance evaluation under varying conversation filtering thresholds, client participation settings, and toxicity thresholds. DeTox-Fed is trained and evaluated with Detoxify-based labels. GraphNLI+RULE uses the same Detoxify-based RULE for training and evaluation; see Section~5.1 for why its scores are high. Bold values indicate the best score within each experimental group and metric column. Results are reported as mean $\pm$ std across random seeds. The number of eligible clients varies according to the dataset and the applied training-size and conversation-length requirements.  All performance metrics are reported in \%.}
\label{tab:detox-fed-detoxify-rule}
\end{table*}

\begin{table*}[h!]
\centering
\resizebox{\textwidth}{!}{%
\begin{tabular}{l HHccc | HHccc}
\toprule

\multirow[b]{2}{*}{\textbf{Method}} &
\multicolumn{5}{c}{\makecell{\textbf{Pleroma} \\ (10 clients, 10 per round)}} &
\multicolumn{5}{c}{\makecell{\textbf{Mastodon} \\ (5 clients, 5 per round)}} \\

\cmidrule(lr){2-6}
\cmidrule(lr){7-11}

&
\textbf{Clients} &
\textbf{\makecell{Clients \\ per Round}} &
\textbf{Macro-F1} &
\textbf{Toxic Precision} &
\textbf{Toxic Recall} &
\textbf{Clients} &
\textbf{\makecell{Clients \\ per Round}} &
\textbf{Macro-F1} &
\textbf{Toxic Precision} &
\textbf{Toxic Recall} \\

\midrule

DeTox-Fed
& 10 & 10
& \fscore{0.7005}{0.0105}
& \fscore{0.6552}{0.0129}
& \fscore{0.6753}{0.0157}
& 5 & 5
& \bfscore{0.7837}{0.0195}
& \bfscore{0.7509}{0.0319}
& \fscore{0.7143}{0.0311}
\\

GraphNLI + \textit{RULE}
& 10 & 10
& \fscore{0.6459}{0.0105}
& \fscore{0.6113}{0.0062}
& \bfscore{0.9653}{0.0031}
& 5 & 5
& \fscore{0.6916}{0.0027}
& \fscore{0.5329}{0.0022}
& \bfscore{0.7972}{0.0092}
\\

GraphNLI + \textit{Calibrated RULE}
& 10 & 10
& \bfscore{0.7450}{0.0046}
& \bfscore{0.7686}{0.0232}
& \fscore{0.7044}{0.0301}
& 5 & 5
& \fscore{0.7412}{0.0011}
& \fscore{0.6485}{0.0026}
& \fscore{0.6616}{0.0087}
\\

\bottomrule
\end{tabular}
}

\caption{
DeTox-Fed vs. GraphNLI performance evaluation with GPT-derived labels.
Results are reported as mean $\pm$ std, with the highest mean value for each metric shown in bold. All performance metrics are reported in \%.
DeTox-Fed setup: Pleroma: $local\_epochs=3$, $learning\_rate=10^{-4}$. Mastodon: $local\_epochs=10$, $learning\_rate=10^{-3}$.}
\label{tab:detox-fed-vs-graphnli-gpt-labels}
\end{table*}

\begin{table*}[ht!]
\centering
\small
\resizebox{\textwidth}{!}{%
\begin{tabular}{lllH ccc ccc}
\toprule

\multirow{2}{*}{\textbf{Model}} &
\multirow{2}{*}{\textbf{Prompt}} &
\multirow{2}{*}{\textbf{Test Set}} &
\multirow{2}{*}{\textbf{Clients}} &
\multicolumn{3}{c}{\textbf{Pleroma}} &
\multicolumn{3}{c}{\textbf{Mastodon}} \\
\cmidrule(lr){5-7}
\cmidrule(lr){8-10}
& & & &
\textbf{Macro-F1} &
\textbf{Toxic Precision} &
\textbf{Toxic Recall} &
\textbf{Macro-F1} &
\textbf{Toxic Precision} &
\textbf{Toxic Recall} \\
\midrule
\multirow{4}{*}{Claude Opus 5}
& \multirow{2}{*}{Base}
& Local
& 10
& \fscore{0.6620}{0.0009}
& \fscore{0.7735}{0.0210}
& \fscore{0.4913}{0.0197}
& \fscore{0.7569}{0.0185}
& \fscore{0.7630}{0.0069}
& \fscore{0.7453}{0.0422}
\\
& & Shared Global
& 10
& \fscore{0.6619}{0.0109}
& \fscore{0.7219}{0.0415}
& \fscore{0.5513}{0.0618}
& \fscore{0.7416}{0.0498}
& \fscore{0.7596}{0.0205}
& \fscore{0.7073}{0.1090}
\\
& \multirow{2}{*}{\textit{RULE}}
& Local
& 10
& \fscore{0.6261}{0.0137}
& \fscore{0.7650}{0.0154}
& \fscore{0.4207}{0.0369}
& \fscore{0.7350}{0.0217}
& \fscore{0.7754}{0.0233}
& \fscore{0.6653}{0.0280}
\\
& & Shared Global
& 10
& \fscore{0.6551}{0.0209}
& \fscore{0.7235}{0.0395}
& \fscore{0.5327}{0.0862}
& \fscore{0.7138}{0.0490}
& \fscore{0.7660}{0.0176}
& \fscore{0.6233}{0.1033}
\\
\midrule
\multirow{4}{*}{GPT-5.6 Sol}
& \multirow{2}{*}{Base}
& Local
& 10
& \bfscore{0.7419}{0.0107}
& \fscore{0.8168}{0.0138}
& \fscore{0.6333}{0.0389}
& \fscore{0.7964}{0.0221}
& \fscore{0.7797}{0.0204}
& \fscore{0.8273}{0.0358}
\\
& & Shared Global
& 10
& \fscore{0.7362}{0.0097}
& \fscore{0.7634}{0.0416}
& \bfscore{0.6940}{0.0520}
& \bfscore{0.8064}{0.0624}
& \fscore{0.7845}{0.0503}
& \bfscore{0.8440}{0.0790}
\\
& \multirow{2}{*}{\textit{RULE}}
& Local
& 10
& \fscore{0.6951}{0.0122}
& \fscore{0.8286}{0.0146}
& \fscore{0.5197}{0.0350}
& \fscore{0.7855}{0.0217}
& \fscore{0.8082}{0.0122}
& \fscore{0.7492}{0.0382}
\\
& & Shared Global
& 10
& \fscore{0.6962}{0.0150}
& \fscore{0.7583}{0.0522}
& \fscore{0.5952}{0.0547}
& \fscore{0.7777}{0.0525}
& \fscore{0.8069}{0.0392}
& \fscore{0.7304}{0.0865}
\\
\midrule
\multirow{4}{*}{Gemini 3.7 Flash}
& \multirow{2}{*}{Base}
& Local
& 10
& \fscore{0.7037}{0.0030}
& \bfscore{0.8433}{0.0177}
& \fscore{0.5264}{0.0126}
& \fscore{0.7387}{0.0077}
& \fscore{0.8492}{0.0176}
& \fscore{0.5953}{0.0160}
\\
& & Shared Global
& 10
& \fscore{0.7029}{0.0102}
& \fscore{0.7825}{0.0573}
& \fscore{0.5843}{0.0644}
& \fscore{0.7241}{0.0402}
& \fscore{0.8388}{0.0661}
& \fscore{0.5733}{0.0375}
\\
& \multirow{2}{*}{\textit{RULE}}
& Local
& 10
& \fscore{0.6665}{0.0169}
& \fscore{0.8415}{0.0286}
& \fscore{0.4536}{0.0184}
& \fscore{0.7143}{0.0171}
& \bfscore{0.8533}{0.0265}
& \fscore{0.5413}{0.0272}
\\
& & Shared Global
& 10
& \fscore{0.6617}{0.0153}
& \fscore{0.7724}{0.0687}
& \fscore{0.5016}{0.0788}
& \fscore{0.6826}{0.0276}
& \fscore{0.8483}{0.0603}
& \fscore{0.4820}{0.0209}
\\
\toprule
\bottomrule

\multirow{2}{*}{Dolphin 3 (8B)$^*$}
& \multirow{2}{*}{Base}
& Local
& 10
& \fscore{0.6517}{0.0158}
& \fscore{0.6427}{0.0012}
& \fscore{0.6893}{0.0775}
& \fscore{0.6127}{0.0245}
& \fscore{0.5914}{0.0157}
& \bfscore{0.8680}{0.0164}
\\
&                         
&                         
Shared Global
& 10
& \fscore{0.6834}{0.0531}
& \fscore{0.6614}{0.0380}
& \bfscore{0.7600}{0.1217}
& \fscore{0.6578}{0.0285}
& \fscore{0.6234}{0.0179}
& \fscore{0.8527}{0.0609}
\\
\midrule
\multirow{2}{*}{Claude Haiku 4.5}
& \multirow{2}{*}{Base}
& Local
& 10
& \fscore{0.6287}{0.0134}
& \bfscore{0.7935}{0.0344}
& \fscore{0.4107}{0.0479}
& \fscore{0.7029}{0.0242}
& \fscore{0.6993}{0.0214}
& \fscore{0.7120}{0.0334}
\\
&                         
&                         
Shared Global
& 10
& \fscore{0.6551}{0.0528}
& \fscore{0.7398}{0.0602}
& \fscore{0.5133}{0.1181}
& \fscore{0.7258}{0.0357}
& \fscore{0.7317}{0.0065}
& \fscore{0.7147}{0.0966}
\\
\midrule
\multirow{2}{*}{GPT-5.4 mini}
& \multirow{2}{*}{Base}
& Local
& 10
& \fscore{0.6956}{0.0083}
& \fscore{0.7566}{0.0221}
& \fscore{0.5913}{0.0508}
& \fscore{0.7304}{0.0360}
& \fscore{0.6954}{0.0352}
& \fscore{0.8333}{0.0300}
\\
&                         
&                         
Shared Global
& 10
& \bfscore{0.7205}{0.0200}
& \fscore{0.7406}{0.0572}
& \fscore{0.6940}{0.0537}
& \bfscore{0.7368}{0.0298}
& \fscore{0.7060}{0.0322}
& \fscore{0.8220}{0.0433}
\\
\midrule
\multirow{2}{*}{Gemini 3.1 Flash-Lite}
& \multirow{2}{*}{Base}
& Local
& 10
& \fscore{0.6609}{0.0196}
& \fscore{0.7704}{0.0123}
& \fscore{0.4922}{0.0560}
& \fscore{0.7152}{0.0196}
& \fscore{0.7562}{0.0257}
& \fscore{0.6413}{0.0363}
\\
&                         
&                         
Shared Global
& 10
& \fscore{0.6744}{0.0280}
& \fscore{0.7173}{0.0645}
& \fscore{0.6015}{0.0898}
& \fscore{0.7153}{0.0719}
& \bfscore{0.7629}{0.0772}
& \fscore{0.6293}{0.0829}
\\
\bottomrule
\end{tabular}
}
\caption{Top-tier vs. lightweight LLM-based moderation using per-client few-shot prompting with Detoxify-derived labels. Results are reported on local and shared global test sets as mean $\pm$ std. Top-tier models are evaluated with both base and \textit{RULE} prompts. Bold values indicate the best score among top-tier and lightweight models for each metric within each dataset. All metrics are reported in \%. *This model was deployed locally through Ollama on Google Colab and, unlike the hosted models, which returned only the label, this model returned a label and reasoning.}
\label{tab:llm_results}
\end{table*}

\begin{table*}[th!]
\centering
\small
\resizebox{\textwidth}{!}{
\begin{tabular}{cccc|ccc|ccc}
\toprule
\multicolumn{4}{c|}{\textbf{Conversation Features}}
& \multicolumn{3}{c|}{\textbf{Pleroma}}
& \multicolumn{3}{c}{\textbf{Mastodon}} \\
\cmidrule(lr){1-4}
\cmidrule(lr){5-7}
\cmidrule(lr){8-10}
\textbf{\makecell{Tree-related\\(DeepWalk)}} &
\textbf{\makecell{User-\\related}} &
\textbf{\makecell{Sentiment-\\based}} &
\textbf{\makecell{Conversation\\Statistics}} &
\textbf{Macro-F1} &
\textbf{Toxic Precision} &
\textbf{Toxic Recall} &
{\textbf{Macro-F1}} &
{\textbf{Toxic Precision}} &
{\textbf{Toxic Recall}} \\
\midrule
\checkmark & -- & -- & --
& \fscore{0.6121}{0.001}
& \fscore{0.6644}{0.003}
& \fscore{0.5186}{0.006}
& {\fscore{0.7050}{0.0085}}
& {\fscore{0.8094}{0.0029}}
& {\fscore{0.7230}{0.0184}}
\\
\checkmark & \checkmark & -- & --
& \fscore{0.6163}{0.000}
& \fscore{0.6683}{0.002}
& \fscore{0.5248}{0.004}
& {\fscore{0.7198}{0.0131}}
& {\fscore{0.8190}{0.0136}}
& {\fscore{0.7404}{0.0214}}
\\
\checkmark & \checkmark & \checkmark & --
& \fscore{0.6959}{0.000}
& \fscore{0.7006}{0.001}
& \fscore{0.7257}{0.005}
& {\fscore{0.7426}{0.0181}}
& {\bfscore{0.8351}{0.0137}}
& {\fscore{0.7623}{0.0256}}
\\
-- & \checkmark & \checkmark & \checkmark
& \bfscore{0.7010}{0.004}
& \fscore{0.6988}{0.003}
& \bfscore{0.7503}{0.010}
& {\fscore{0.6877}{0.0130}}
& {\fscore{0.7992}{0.0102}}
& {\fscore{0.7005}{0.0195}}
\\
\checkmark & \checkmark & \checkmark & \checkmark
& \fscore{0.6950}{0.001}
& \bfscore{0.7007}{0.000}
& \fscore{0.7220}{0.006}
& {\bfscore{0.7463}{0.0187}}
& {\fscore{0.8343}{0.0116}}
& {\bfscore{0.7720}{0.0299}}
\\
\bottomrule
\end{tabular}
}
\caption{
DeTox-Fed feature ablation study. The evaluated feature groups are described in Section~\ref{sec:features}. Results are reported as mean $\pm$ std, with the best mean value for each metric shown in bold. All metrics are reported in \%.}
\label{tab:feature_ablation_comparison}
\end{table*}

\noindent\textbf{Cost and Time.}
We trained and tested both models on the Mastodon dataset under the GPT-based evaluation setting to compare their computational efficiency (Table ~\ref{tab:training_inference_time}). DeTox-Fed and CPU GraphNLI were executed on a 64-core Linux HPC server ($\sim$1~TB RAM), while GPU GraphNLI used 4$\times$ NVIDIA RTX 6000 Ada GPUs (48~GB each, 512~GB RAM). Combined training and inference costs for both models are estimated using the closest AWS on-demand equivalents, \texttt{x2idn.16xlarge} (\$6.67/h) and \texttt{g6e.12xlarge} (\$10.49/h). DeTox-Fed required $\sim$3~min training and 0.46~s inference ($\sim$\$0.32), compared with GraphNLI's 76.8~h/12~min on CPU ($\sim$\$513) and 1.75~h/28.8~s on GPU ($\sim$\$18.4), demonstrating DeTox-Fed's substantially lower computational cost.

\subsection{DeTox-Fed vs. LLMs}
\label{sec:detoxfed-vs-llms}

\noindent\textbf{Top-tier models.}
As shown in Table~\ref{tab:llm_results}, GPT-5.6 Sol achieves the highest performance among the top-tier models under the base prompt, reaching 74.19\% Macro-F1 on the Pleroma local set and 80.64\% on the Mastodon global set. In contrast to the lightweight models, the top-tier models do not consistently perform better on the global test set, and their results show noticeable variation across the evaluated settings. Moreover, the higher cost of top-tier models does not consistently translate into better performance. For example, Claude Opus 5 performs below the substantially cheaper GPT-5.4 mini across the Pleroma configurations.

\noindent\textbf{Lightweight models.}
All lightweight models score higher on the shared global test set than on their local equivalents. GPT-5.4 mini performs best overall, achieving a Macro-F1 of 72.05\% on Pleroma and 73.68\% on Mastodon. While hosted models generally prioritize precision over recall on Pleroma, Dolphin 3 maximizes toxic recall across both platforms (86.80\% on Mastodon, 76\% on Pleroma). In a semi-automated moderation pipeline where models serve as an initial filter, this high recall is crucial, as a missed toxic conversation is costlier to the community than a false positive.

\noindent\textbf{Effect of the \textit{RULE}-Aware Prompt.}
We also evaluated the \textit{RULE}-aware prompt that explicitly describes the labeling \textit{RULE}. This consistently reduced Macro-F1 across all 12 comparisons, mainly due to lower recall. The results suggest that explicitly prompting LLMs to apply numerical reply-count and fraction thresholds makes them more conservative and does not improve classification performance.

\noindent\textbf{Cost and Time.}

API-based LLMs do not require local model training, so their reported cost corresponds to inference only, whereas the DeTox-Fed cost reported in Table~\ref{tab:training_inference_time} includes both training and inference. The evaluated LLM runs cost approximately \$5--\$48 on Pleroma and \$7--\$53 on Mastodon, depending on the model and prompt (see Table~\ref{tab:llm-cost} in the Appendix). Runtime is also affected by provider rate limits and batch-processing queues. These results highlight a key deployment trade-off: DeTox-Fed requires an initial FL training stage but supports inexpensive local inference, whereas API-based LLM moderation incurs recurring inference costs. This difference becomes increasingly important at scale: once trained, DeTox-Fed performs local inference without per-query API costs or provider rate limits, whereas hosted LLM moderation remains subject to these overheads for each new conversation.

\subsection{DeTox-Fed Sensitivity and Ablation Analysis}

\noindent\textbf{A) Sensitivity Analysis.}
We evaluate DeTox-Fed with Detoxify-based labels through controlled experiments testing its robustness and sensitivity to moderation settings.

On Pleroma, Macro-F1 remains relatively stable around 68-70\% across most configurations, while toxic-class precision and recall vary more substantially. On Mastodon, performance is more sensitive to training size and conversation length, while remaining relatively stable across toxicity thresholds.

\noindent\textbf{Effect of Training Size.}
We examine how performance changes as the number of labeled training conversations available to each client varies. This experiment reflects realistic moderation settings, where only a limited number of conversations may be manually labeled by instance moderators. Increasing the number of labeled conversations improves performance on both datasets. On Pleroma, Macro-F1 largely saturates around 200 labels per client, while toxic precision increases from 67.28\% to 70.50\% and recall from 63.10\% to 72.49\%. Mastodon benefits more strongly from additional supervision: Macro-F1 increases from 63.11\% to 74.55\%, precision from 75.44\% to 83.61\%, and recall from 66.61\% to 76.68\%. This suggests that additional supervision mainly helps the model identify toxic conversations more reliably.

\noindent\textbf{Effect of Conversation Length.}
We study the effect of conversation-length filtering by varying the minimum number of posts required for a conversation to be included. This allows us to assess whether longer conversations provide richer structural and interaction signals for toxicity detection. On Pleroma, filtering very short conversations improves Macro-F1, with the best result obtained at five posts. A stricter threshold of 10 posts substantially increases toxic precision, from 53.69\% to 82.85\%, while recall changes slightly and Macro-F1 decreases to 67.60\%. This means that longer conversations make toxic cases easier to detect, but the model becomes less balanced across both classes,  probably because the filtered dataset is smaller and less representative. On Mastodon, longer conversations consistently improve all metrics, reaching 74.51\% Macro-F1, 83.55\% precision, and 76.73\% recall at ten posts.

\noindent\textbf{Effect of Client Number Per Round.}
We evaluate robustness to partial client participation by varying the number of clients selected in each FL round. Changing the number of participating clients per FL round has little effect on Pleroma: Macro-F1 remains around 69.5\%, with similarly stable toxic precision
and recall even with only 10 clients per round. Therefore, the method appears robust to partial client participation, which is realistic for decentralized social networks. We do not perform the same sensitivity analysis on Mastodon because the selected configuration contains only a small number of eligible instances.

\noindent\textbf{Effect of Moderator's Tolerance to Toxicity.}
We analyze the effect of the toxicity threshold used to generate conversation-level labels, capturing moderation settings from stricter to more permissive definitions of toxicity. On Pleroma, increasing the threshold gradually reduces Macro-F1 and toxic precision, while toxic recall remains relatively stable. This suggests that DeTox-Fed can adapt to different moderation settings, although more permissive thresholds make precise toxic-class detection harder. On Mastodon, Macro-F1 remains stable around 74-75\%, while toxic precision decreases from 86.46\% to 74.70\% and recall from 78.46\% to 74.19\%. Overall, the threshold affects class-specific precision and recall more strongly than
balanced performance.

Finally, we conduct a feature ablation study to assess the contribution of structural, social, sentiment-based, and conversation-level features to the final classification performance.

\noindent\textbf{B) Feature Ablation Study.}
Sentiment features provide the largest improvement on both datasets. On Pleroma, Macro-F1 increases from 61.63\% to 69.59\% when sentiment is added, while the best result (70.10\%) is obtained without DeepWalk.
In contrast, Mastodon performs best with the complete feature set (74.63\%), while removing DeepWalk reduces Macro-F1 to 68.77\%. Importantly, DeepWalk captures the internal structure of individual conversation trees and is distinct from the conversation graph on which the GNN performs message passing. Thus, the Pleroma result does not imply that graph-based learning is unnecessary, but rather that tree-level structural embeddings provide limited additional information for its conversations. Their stronger effect on Mastodon further suggests that their usefulness is dataset-dependent.

\section{Discussion}
\label{sec:discussion}

\noindent\textbf{Framework implications and comparison with LLMs.}
Our results show that DeTox-Fed can operate under limited local supervision and partial client participation, while the usefulness of individual conversation-level feature groups varies across datasets. We do not claim that DeTox-Fed consistently outperforms text-only LLMs. The few-shot LLMs achieve comparable performance and, in some settings, higher toxic-class recall, which is particularly relevant when moderation systems are used as an initial human-in-the-loop filter.
Instead, the practical contribution of DeTox-Fed lies in providing a conversation-level moderation framework that combines relational,
conversation, user, and sentiment information while supporting collaborative learning without exchanging raw conversations or moderation labels.
Experiments with a small number of clients further show that performance remains generally stable when fewer clients participate in each FL round. 
Nevertheless, the FL and LLM settings are not directly comparable in this respect, since the LLM classifiers are not trained collaboratively across clients.

\noindent\textbf{Deployment, data-locality and cost trade-offs.}
DeTox-Fed and LLMs also represent different deployment models. DeTox-Fed keeps raw conversations and moderation labels local, although model updates are still shared, and FL alone does not provide a formal privacy guarantee. Locally hosted LLMs avoid third-party data exposure, but may require substantially greater local computational resources. Overall, DeTox-Fed provides a substantially cheaper deployment option than hosted LLMs. Normalized per 1,000 conversations, hosted LLM inference costs approximately \$1.09--\$8.82, compared with \$0.13 for DeTox-Fed's combined training and inference, making hosted LLMs roughly 8.5--69$\times$ more expensive.

The LLM experiments also expose practical limitations of the current hosted moderation services. Few-shot performance may depend on the selected examples, and safety mechanisms can occasionally refuse to process severe toxic content. For example, Gemini 3.1 Flash-Lite refused $0.23\%$ of our requests. Provider-side behavior can also affect cost independently of the requested output, as observed for Gemini 3.7 Flash, where billed thinking tokens substantially exceeded the single-word classification output. These effects are particularly relevant when moderation is performed continuously and on a large scale.

\noindent\textbf{Generalization and limitations.}
Although the evaluation was conducted on two datasets, we do not assume that DeTox-Fed's performance will remain unchanged on other Fediverse platforms, such as Lemmygrad. Differences in toxicity prevalence, user activity, conversation structure, and moderation norms may affect both class distributions and feature utility. Evaluation on additional platforms and communities, therefore, remains necessary.

Our current graphs are also static snapshots. Because conversation edges are derived from shared-user participation, user attrition, new-user uptake, and changing participation patterns may alter the graph over time. Evaluating temporal robustness will require reconstructing conversation graphs across multiple periods and measuring performance under evolving co-participation patterns.

The current evaluation also relies primarily on scalable synthetic annotation rather than large-scale human-validated conversation labels. In particular, the LLM comparison using Detoxify-derived reference labels
primarily measures agreement with that annotation scheme rather than
human-grounded toxicity accuracy. Future work should therefore evaluate the
framework using larger, independently human-annotated datasets and
instance-specific moderation policies. Finally, binary toxic/non-toxic classification simplifies real moderation, which may require distinctions between harassment, hate speech, threats, spam, and other policy violations.

\noindent\textbf{Extensions.}
A natural extension of DeTox-Fed is to move beyond local instance-level conversation graphs. Conversations hosted on different Fediverse instances can be connected through shared-user participation, allowing local graphs to be selectively expanded with one- or two-hop neighboring conversations. Such inter-instance expansion could provide additional conversational context, but its communication overhead, scalability, and privacy implications require dedicated evaluation. More broadly, the federated conversation-graph formulation is not inherently limited to toxicity detection and could support other conversation-level moderation and safety tasks.

\section{Related Work}
\label{sec:related_work}

\subsection{Content Moderation in Decentralized Social Networks}
Recent work has focused on content moderation and governance in DSNs. \citet{agarwal2024decentralised} propose GraphNLI, a decentralized, conversation-aware moderation approach for Pleroma and the Fediverse, showing that individual servers observe only a partial view of federated conversations and that moderation must therefore operate under local visibility constraints.
Unlike GraphNLI, which exploits the structure within individual conversations,
and prior graph-based moderation methods that model interactions on centralized
platforms, DeTox-Fed constructs a graph of conversations linked through shared
user participation and learns over these graphs in a federated setting.

Collaborative moderation for toot-level toxicity classification has also been explored in DSNs, motivated by the limited labeled data available at individual DSN instances. MODPAIR~\cite{ModPair_pleroma_data} studies toxicity detection in the Fediverse using the Pleroma dataset and proposes decentralized model sharing between instances through the exchange of locally trained model parameters. However, the approach relies on text-based classifiers such as Logistic Regression and Support Vector Machines and does not explicitly model conversation-level relational context. Their results further show that smaller DSN instances benefit significantly from collaborative model sharing.

Recent research on community-level blocklists further shows that decentralized moderation depends on inter-instance governance mechanisms, where communities decide which other communities to block or trust~\cite{zhang2025blocklists}. Broader work on governance in online communities also supports this view. \citet{leibmann2025reddit} study rules across thousands of Reddit communities and show that community rules, their wording, and their evolution over time relate to users' perceptions of governance. Although this work focuses on Reddit rather than DSNs, it reinforces the idea that moderation depends strongly on community-specific norms. 

Federated and decentralized learning have also started to appear in other Fediverse-related safety tasks beyond toxicity detection. FediScan~\cite{gao2026fediscan} proposes a decentralized federated learning framework for social bot detection in the Fediverse. Although it focuses on bot detection rather than toxic conversation detection, it addresses a similar deployment challenge: building safety models when data and governance remain distributed across independent instances.

\subsection{Content Moderation in Mainstream Platforms}
Several works study harmful content detection on mainstream, non-decentralized platforms. HypergraphDis~\cite{salamanos2024hypergraphdis} models online disinformation detection using hypergraph-based representations, capturing higher-order relationships that go beyond pairwise interactions. Similarly, Meta-graph~\cite{metagraph} uses graph-based representations to model complex interaction structures for harmful or misleading content detection. These works show that graph-based modeling can improve moderation and disinformation detection by capturing relational context that text-only models may miss. However, they focus on mainstream centralized settings, while our work addresses DSNs, where data ownership, moderation rules, and infrastructure remain distributed across independent instances. 

\subsection{LLMs for Toxicity Detection and Moderation}
Recent work has examined LLMs as tools for content moderation and toxicity detection. \citet{kumar2024watch} evaluate publicly available LLMs on rule-based community moderation and toxic content detection, showing that LLMs can support moderation tasks but remain sensitive to community-specific norms and task settings. \citet{li2024promise} study ChatGPT for detecting hateful, offensive, and toxic comments on social media, highlighting the potential of prompt-based LLM classifiers.

Other approaches use LLM prompting to improve toxicity detection pipelines. \citet{zhang2023efficient} use LLM reasoning to improve toxic content detection, while \citet{hu2024toxicity} examine LLM refusal behavior as a signal for identifying toxic prompts. \citet{deshpande2023toxicity} further show that LLM behavior around toxicity can change depending on persona-based prompting. These studies suggest that LLMs can provide useful moderation signals, but their performance depends strongly on prompting strategy, model alignment, and the type of toxic content.

\section*{Conclusion}
\label{conclusion}
We presented DeTox-Fed, a federated graph-learning framework for
conversation-level toxicity detection in DSNs. The framework models each instance as a local conversation graph, where nodes
correspond to conversation trees and edges capture shared user participation across conversations. By training a GNN in an FL setup, DeTox-Fed allows instances to collaboratively learn a toxicity classifier without sharing raw conversations, moderation labels, or user-level data. Experiments on Pleroma and Mastodon show that DeTox-Fed remains effective under limited local supervision and different moderation settings, while the Pleroma experiments further demonstrate robustness to partial client participation. The comparison with GraphNLI and LLM-based approaches shows that no method consistently dominates across datasets. DeTox-Fed remains competitive while providing substantially lower computational cost than GraphNLI. From a deployment cost perspective, DeTox-Fed is therefore substantially cheaper than both GraphNLI and hosted LLM-based moderation. These findings suggest that federated graph-based learning is a promising direction for semi-automated moderation in DSNs, especially in settings where communities need to retain local control over their data and moderation
policies.

\section*{Ethics}
This work followed established principles and guidelines for ethical
information research and the use of shared data \cite{ethics}.
For Pleroma, we used an existing published dataset. For Mastodon, we collected publicly accessible posts and replies from 20 instances through the official Mastodon REST API. Our analysis does not report usernames, account handles, or other directly identifying user information. Results are presented only in aggregate form. The data collection and analysis were designed to minimize the exposure of personal information and to respect applicable privacy and data-protection principles.

\section*{Acknowledgments}

This work has been funded by the European Union under the projects: MedDMO II (Grant Agreement no. 101226175), CYberSafety V (Grant Agreement no. 101158509), and EXARCH (Grant Agreement no. 101236244).

\fontsize{9pt}{10pt}\selectfont
\bibliography{Paper}

\newpage
\appendix
\section*{APPENDIX}
\label{sec:appendix}

\begin{figure}[h!]
    \centering

    \begin{subfigure}{0.72\columnwidth}
        \centering
        \includegraphics[width=\linewidth]
        {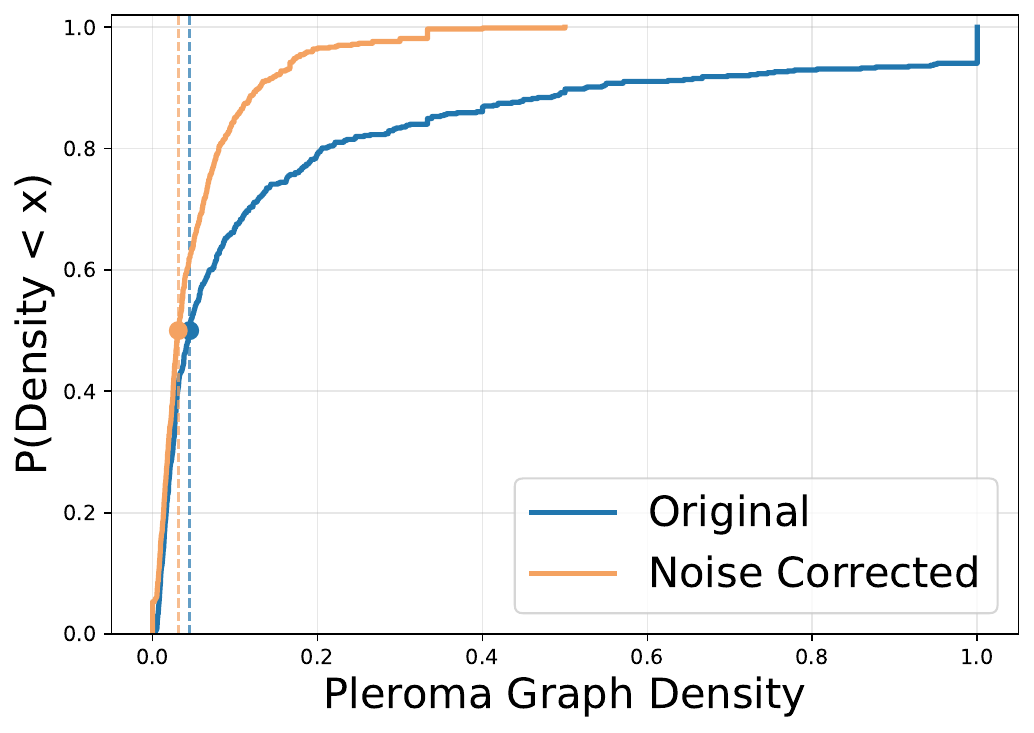}
        \caption{}
    \end{subfigure}
    \hfill
    \begin{subfigure}{0.72\columnwidth}
        \centering
        \includegraphics[width=\linewidth]
        {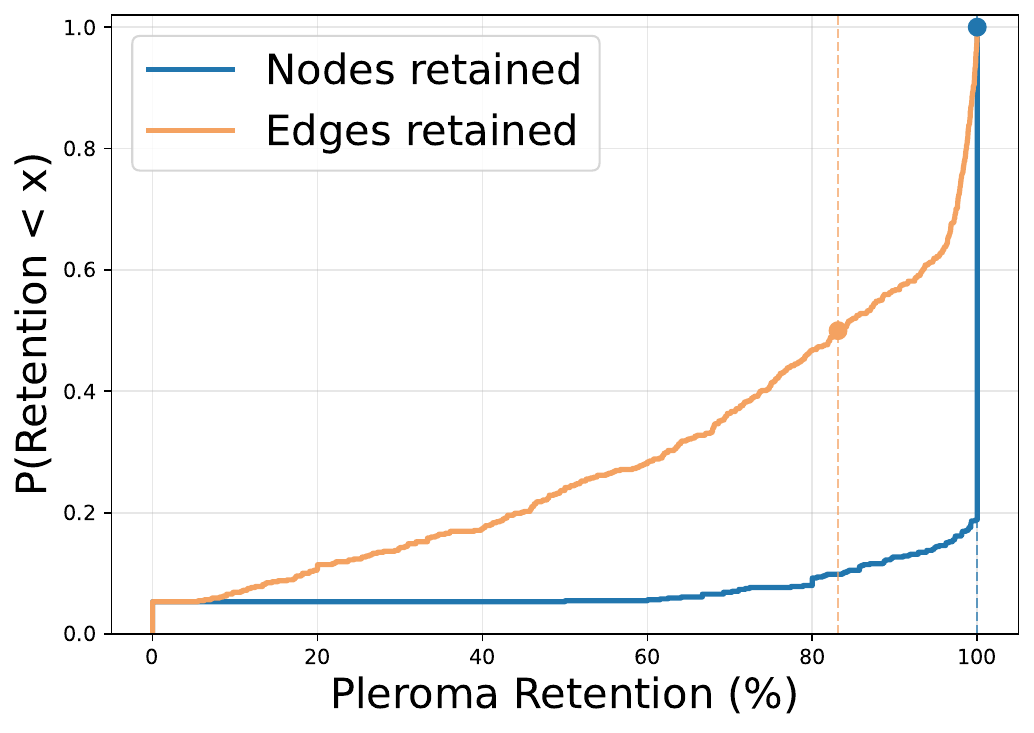}
        \caption{}
    \end{subfigure}

    \vspace{0.3em}

    \begin{subfigure}{0.72\columnwidth}
        \centering
        \includegraphics[width=\linewidth]
        {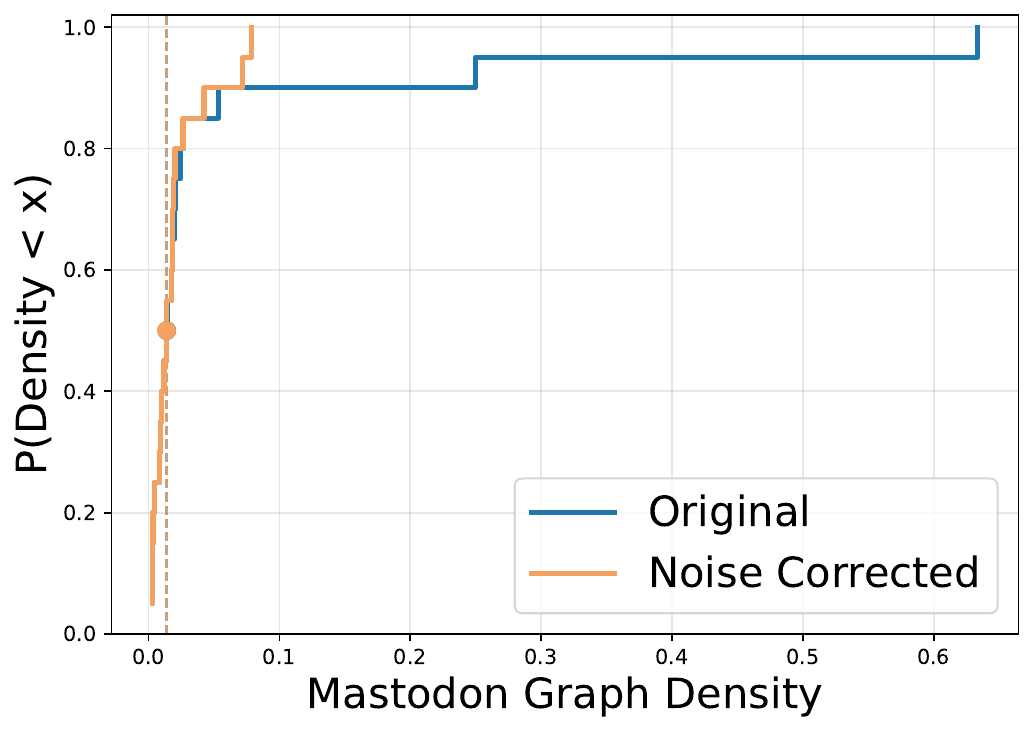}
        \caption{}
    \end{subfigure}
    \hfill
    \begin{subfigure}{0.72\columnwidth}
        \centering
        \includegraphics[width=\linewidth]
        {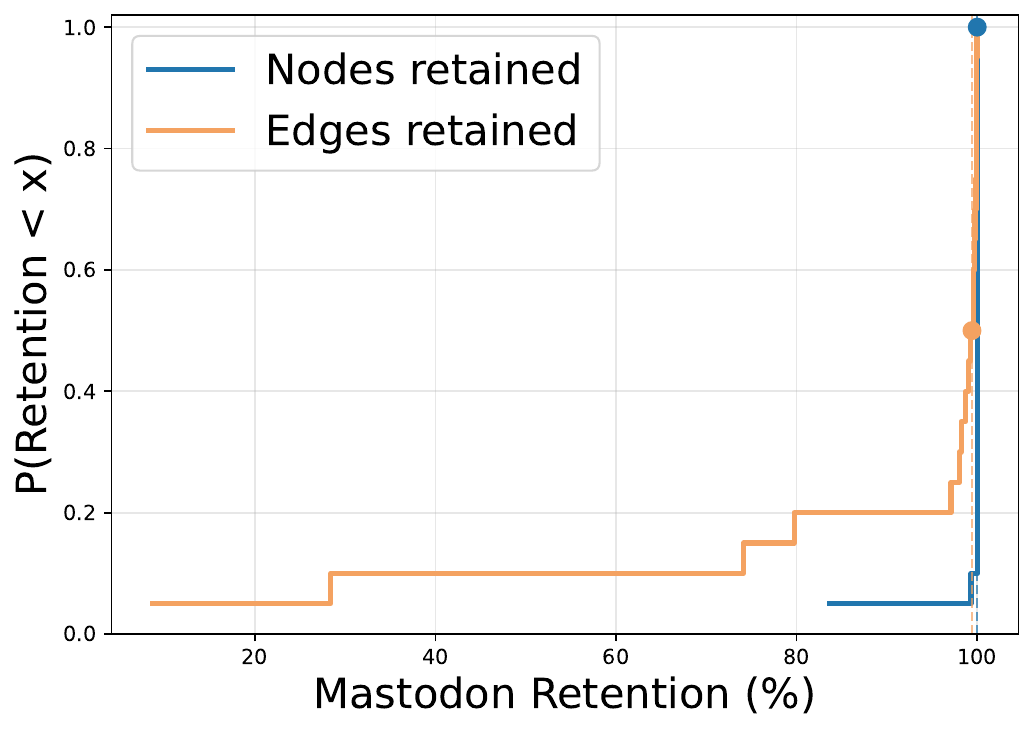}
        \caption{}
    \end{subfigure}

    \caption{Effect of noise-corrected backboning on the conversation
    graph of each instance in Pleroma and Mastodon.
    (a) and (c) Density before and after Noise-Corrected Backboning.
    (b) and (d) Cumulative Distribution Function (CDF)
    of node and edge retention after noise reduction.}
    \label{fig:backbone_effects}
\end{figure}

\begin{table}[t]
\small
\centering
\resizebox{\columnwidth}{!}{%
\begin{tabular}{lcccHc}
\makecell{Model} &
\textbf{\makecell{Cost (\$)}} &
\textbf{Tokens} &
\textbf{\$/1M} &
\textbf{\makecell{\$/\\1KConv.}} &
\textbf{\makecell{Runtime (h)}} \\

\midrule

\multicolumn{6}{c}{\textbf{Pleroma}}\\
\midrule
\midrule
\textbf{\makecell[l]{Claude \\Haiku 4.5}}
& 12.33
& 34.31M
& 0.36
& 2.06
& $\sim$2.5
\\

\makecell[l]{\textbf{GPT-5.4 mini} }
& \textbf{5.38}
& \textbf{30.28M}
& \textbf{0.18}
& \textbf{0.90}
& $\sim$2.5
\\

\makecell[l]{Gemini 3.1 \\Flash-Lite}
& 10.72
& 32.30M
& 0.33
& 1.79
& $\sim$3.0
\\
\midrule

\textbf{Claude Opus 5:}

& 46.83
& 46.84M
& 1.00
& 7.81
& \multirow{2}{*}{$\sim$0.25}
\\
\cmidrule(lr){1-1}
\textit{RULE Prompt}
& 48.44
& 47.38M
& 1.02
& 8.07
&
\\
\midrule
\textbf{GPT-5.6 Sol:} 

& 41.82
& \textbf{30.57M}
& 1.37
& 6.97
& \multirow{2}{*}{$\sim$3.5}
\\

\cmidrule(lr){1-1}
\textit{RULE Prompt}
& 42.41
& 30.91M
& 1.37
& 7.07
&
\\
\midrule
\textbf{\makecell[l]{Gemini 3.7 Flash:}}
& \textbf{11.92}
& 32.73M
& \textbf{0.36}
& \textbf{1.99}
& \multirow{2}{*}{$\sim$0.6}
\\
\cmidrule(lr){1-1}
\textit{RULE Prompt}
& 12.12
& 33.07M
& 0.37
& 2.02
&
\\

\midrule
\midrule

\multicolumn{6}{c}{\textbf{Mastodon}}\\
\midrule
\midrule
\textbf{\makecell[l]{Claude \\Haiku 4.5}}
& 9.32
& 43.25M
& 0.22
& 1.55
& $\sim$2.5
\\

\textbf{GPT-5.4 mini }
& \textbf{6.54}
& \textbf{38.03M}
& \textbf{0.17}
& \textbf{1.09}
& $\sim$2.5
\\

\textbf{\makecell[l]{Gemini 3.1 \\Flash-Lite}}
& 11.80
& 42.45M
& 0.28
& 1.97
& $\sim$4.0
\\
\midrule

\textbf{Claude Opus 5:}

& 43.32
& 61.23M
& 0.71
& 7.22
& \multirow{2}{*}{$\sim$0.25}
\\
\cmidrule(lr){1-1}
\textit{RULE Prompt}
& 40.24
& 61.76M
& 0.65
& 6.71
&
\\
\midrule

\textbf{GPT-5.6 Sol }
& 52.44
& \textbf{38.06M}
& 1.38
& 8.74
& \multirow{2}{*}{$\sim$3.5}
\\
\cmidrule(lr){1-1}

\textit{RULE Prompt}
& 52.92
& 38.38M
& 1.38
& 8.83
&
\\
\midrule

\textbf{Gemini 3.7 Flash}
& \textbf{14.21}
& 40.16M
& \textbf{0.35}
& \textbf{2.37}
& \multirow{2}{*}{$\sim$0.6}
\\
\cmidrule(lr){1-1}
\textit{RULE Prompt}
& 14.52
& 40.50M
& 0.36
& 2.42
&
\\

\bottomrule
\end{tabular}
}
\caption{Cost and runtime of LLM-based few-shot classification across models on the Pleroma and Mastodon datasets, including three top-tier models under both the base and \textit{RULE}-aware prompts. Total tokens include input, cached, and output tokens for the full evaluation run of 6K classifications. Runtime for the lightweight models is the wall clock time of that run under a client-side cap of 50 requests per minute, with one request in flight at a time, so it reflects throughput under our rate limiting rather than the latency of a single call. The three top-tier models were submitted through each provider's batch API. Batch API runtime varies across providers, from about 15 minutes for Opus 5 to about 3.5 hours for GPT-5.6 Sol, and is set by provider-side queuing rather than by our workload. Their runtime instead reflects provider-side batch turnaround and is reported once per model since both prompt variants were submitted together. Bold values indicate the lowest value per metric column, computed separately within the lightweight and top-tier groups in each dataset.}
\label{tab:llm-cost}

\end{table}

\end{document}